\documentclass{aastex631}

\usepackage{placeins}
\usepackage{amsmath}

\begin{document}

\title{BSN: The First Photometric Analysis of Contact Binary Systems V1961 Cyg and V0890 Lyr}

\author[0009-0004-8426-4114]{Sabrina Baudart}
\altaffiliation{sabrina.baudart@gmail.com}
\affiliation{Double Stars Committee, Société Astronomique de France, Paris, France}

\author[0000-0002-0196-9732]{Atila Poro}
\altaffiliation{atilaporo@bsnp.info}
\affiliation{LUX, Observatoire de Paris, CNRS, PSL, 61 Avenue de l'Observatoire, 75014 Paris, France}

\begin{abstract}
We presented the first photometric analysis of the V1961 Cyg and V0890 Lyr binary systems. We observed and analyzed these systems at an observatory in France as part of the Binary Systems of South and North (BSN) Project. We extracted and collected the times of minima from the observations and literature and presented a new ephemeris for each system. Due to the few observations about these systems over the years, both O-C diagrams were fitted linearly. The PHysics Of Eclipsing BinariEs (PHOEBE) Python code and the Markov Chain Monte Carlo (MCMC) method were used to light curve solutions. The light curve solution required a cold starspot on the hotter component in the V1961 Cyg binary system. We compared and have close agreements between our mass ratios' results from the light curve analysis processes and a new method based on the light curve derivative. We estimated the absolute parameters using an empirical relationship between the semi-major axis and orbital period for contact binary systems. The results show V1961 Cyg and V0890 Lyr are W-type contact binary systems. We displayed stars and systems' positions in the $M-L$, $M-R$, and $logM_{tot}-logJ_0$ diagrams. We also presented a new relationship between mass ratio and luminosity ratio.
\end{abstract}

\keywords{binaries: eclipsing - methods: observational - stars: individual (V1961 Cyg and V0890 Lyr)}

\section{Introduction}
Two late-type stars in W Ursae Majoris (W UMa) eclipsing contact binary systems have short orbital periods. The two components overfill their critical Roche lobes in these systems and share a common envelope (\citealt{1959cbs..book.....K}). The effective temperature difference between the component stars in contact binary systems is low, and they additionally have a close or equal depth of minima (\citealt{kuiper1941}, \citealt{2005ApJ...629.1055Y}).

W UMa contact binaries are generally classified into two categories: A-subtype and W-subtype (\citealt{binnendijk1970orbital}). The subtype of a system cannot be recognized only from its light curve shape, and it is required to estimate its absolute parameters, including the mass of the stars (\citealt{guo2022first}).

The stars of contact systems are transferring mass to each other (\citealt{1979ApJ...231..502L}); in this process, their orbital periods can be changed. The orbital period of contact systems plays a role in relations with absolute parameters, and they are effective in the evolutionary process of these systems (\citealt{2021ApJS..254...10L}, \citealt{2022MNRAS.514.5528L}, \citealt{2024NewA..11002227P}, \citealt{2024RAA....24a5002P}). Several studies have been conducted on the upper and lower cut-offs of these systems' orbital periods (\citealt{2020MNRAS.497.3493Z}).

Asymmetric light curves in contact and near-contact binary are commonly observed over time, especially in phases 0.25 and 0.75. This phenomenon is generally referred to as the O'Connell effect (\citealt{1951MNRAS.111..642O}), which is crucial for studying a star's magnetic activity. This asymmetry in the light curves is solved with one or more starspot(s) and is challenging in the modeling.
\\
\\
In this work, we investigated V1961 Cyg (2MASS J21243169+3957197) and V0890 Lyr (2MASS J19184581+3708166) binary star systems classified as EW type in the variable stars' catalogs and databases such as ASAS-SN\footnote{The All Sky Automated Survey, \url{https://asas-sn.osu.edu/variables}}, GCVS\footnote{General Catalogue of Variable Stars}, ZTF\footnote{The Zwicky Transient Facility}, VSX\footnote{International Variable Star indeX, \url{https://www.aavso.org/vsx/}}.

The variability of V1961 Cyg (2MASS J21243169+3957197) was discovered by \cite{kulagin1989new}. Based on the Gaia DR3, V1961 Cyg has coordinates of RA. $321.1320^{\circ}$ and Dec. $39.9554^{\circ}$ in the Cygnus constellation. This system ranges from 13.88 to 14.6 magnitudes in the $V$ filter, according to the VSX database. The orbital period of V1961 Cyg was reported to be 0.286008 days in the \cite{kulagin1989new} study and 0.2860135 days in the ASAS-SN catalog of variable stars.

In the survey of the Two Microns All Sky Survey (2MASS) and the Northern Sky Variability Survey (NSVS) data, V0890 Lyr was found (\citealt{gettel2006catalog}). This system is located in the Lyra constellation with coordinates of RA.$289.6909^{\circ}$ and Dec. $37.1379^{\circ}$, according to the Gaia DR3 database. The VSX database reported a magnitude range of $11.38–11.71$ magnitudes in the $V$ filter and an orbital period of 0.3884152 days for V0890 Lyr.
\\
\\
We aim to investigate two contact binary systems that have not been previously studied. Different systems have their characteristics, and their analysis can lead to a better understanding of them. Also, investigating new contact binary systems might result in a larger sample of the studied targets, which could be used to study future empirical parameters. This paper's structure is as follows: Section 2 describes the observational data and methods used for data reduction. Section 3 presents the new ephemeris of each contact binary system. In Section 4, we present the solution of the light curve analysis, while Section 5 gives the estimated absolute parameters of each target system. Finally, Section 6 gives our discussion and conclusion.

\vspace{1cm}
\section{Observation and Data Reduction}
We have observed V1961 Cyg and V0890 Lyr at a private observatory located in Toulon, France (longitude $05^\circ$ $54'$ $35"$ E, latitude $43^\circ$ $8'$ $59"$ N, altitude 68 meters above sea level). The photometric observations were made using the standard $V$ filter for both target systems. We employed an apochromatic refractor telescope with a 102mm aperture and a ZWO ASI 1600MM CCD to observe the V1961 Cyg and V0890 Lyr binary systems. The binning of the images was $1\times1$, and the average temperature during the observations was $-15^\circ$C.

\begin{figure*}
    \centering
    \includegraphics[width=\textwidth]{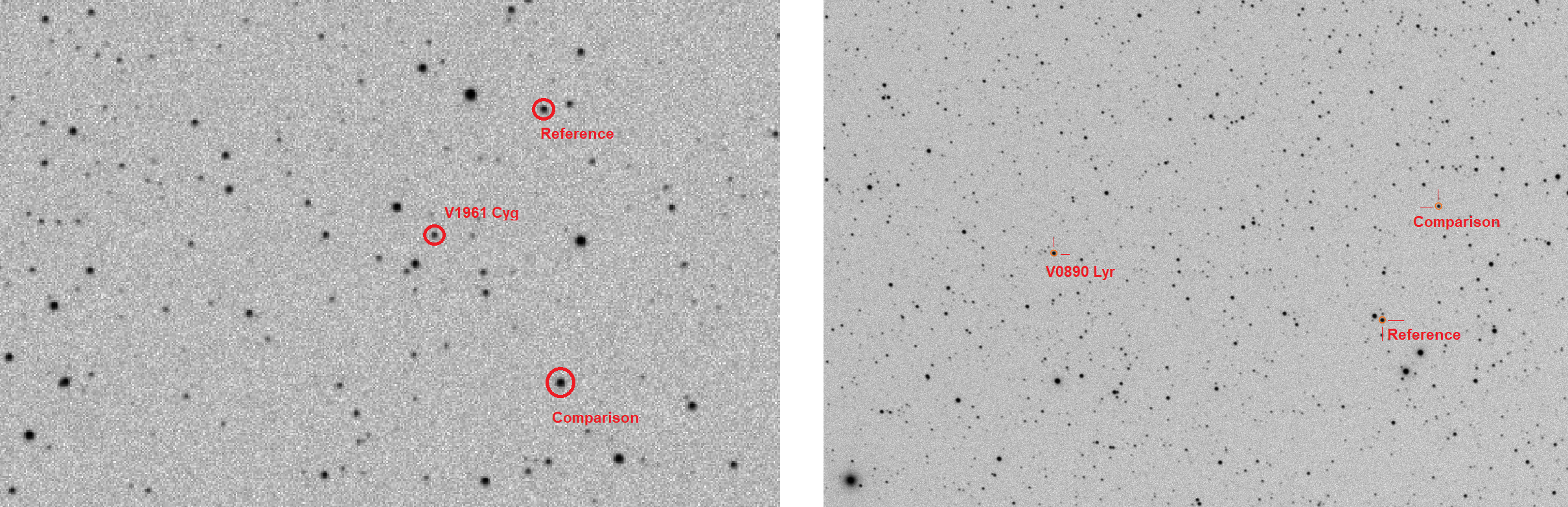}
    \caption{The left side is the field-of-view of the V1961 Cyg binary system, and the right side is for V0890 Lyr, together with the comparison and check stars used in the observations.}
    \label{Observations}
\end{figure*}

V1961 Cyg was observed for three nights on July 2nd, 9th, and 13th, 2023, accumulating 499 images with 110 seconds of exposure time. We used Gaia DR2 1965471833170939904, whose magnitude is $V=12.795(28)^{mag.}$ and coordinates are RA.: $21^h$ $24^m$ $21.976^s$, Dec.: $+39^\circ$ $59'$ $36.83"$, as a comparison star. Gaia DR2 1965424588530688256 was chosen as a check star with a magnitude of $V=13.621(34)^{mag}$ and coordinates RA: $21^h$ $24^m$ $22.725^s$, Dec.: $+39^\circ$ $55'$ $27.37"$ (Figure \ref{Observations} left). Our observations' average Signal-to-Noise Ratio (SNR) is 15.49 decibels (dB), which indicates a 0.03 magnitude of uncertainty for V1961 Cyg.

V0890 Lyr was observed for two nights on August 19th and 20th, 2023. These observations resulted in 484 images with an exposure time of 80 seconds. Gaia DR2 2051151521582207232 (RA: $19^h$ $17^m$ $06.764^s$, Dec.: $+37^\circ$ $06'$ $03.74"$,$V=12.672(140)^{mag}$ ) was used as a check star, and Gaia DR2 2051169835322822912 (RA: $19^h$ $17^m$ $21.475^s$, Dec.: $+37^\circ$ $11'$ $52.05"$,$V=11.439(119)^{mag}$) as a comparison star (Figure \ref{Observations} right). Our data analysis indicates that the average SNR is 24.56 dB, which means an uncertainty of 0.004 magnitude for V0890 Lyr.

The field-of-view of the target systems with comparison and check stars is shown in Figure \ref{Observations}.

We obtained each light curve's apparent magnitude of maximum brightness based on our observations. We resulted in $V_{max}=13.83(37)$ magnitude for V1961 Cyg and $V_{max}=11.627(90)^{mag}$ for V0890 Lyr.

The data reduction process for both binary systems involved standard calibration techniques, including the correction for bias using offset frames, subtraction of dark frames to eliminate thermal noise, and division by flat-field frames to correct for uneven illumination. These standard processes were executed using the 1.2.0 version of the Siril software\footnote{\url{https://siril.org/fr/download/2023-09-15-siril-1.2.0/}}.

\vspace{1cm}
\section{New Ephemeris}
Data derived from our ground-based observations were used to extract primary and secondary times of minima. We use a Python program to fit a Gaussian distribution and extract minima from ground-based data (Table \ref{tab1}). We used the online tool\footnote{\url{https://astroutils.astronomy.osu.edu/time/hjd2bjd.html}} to convert the times of minima gathered in the literature, which were Heliocentric Julian Day (HJD), to Barycentric Julian Date and Barycentric Dynamical Time ($BJD_{TDB}$). The times of minima are listed in Table \ref{tab1}.

\begin{table*}
\caption{Available CCD times of minima from this study and literature for V1961 Cyg and V0890 Lyr binary systems.}
\centering
\begin{center}
\footnotesize
\begin{tabular}{c | c c c c c}
 \hline
 \hline
System & Min.($BJD_{TDB}$) & Error & Epoch & O-C & Reference\\
\hline
 & 2457891.08346	& & 16194 & 0.0003 & ASAS-SN \\
 & 2454291.60675	& & 3609 & 0.0035 & VSX \\
 V1961 Cyg & 2453259.38056	& 0.00240 & 0 & 0.0000 & \cite{2005IBVS.5657....1H} \\
 & 2453259.52356	& 0.00280 & 0.5	& 0.0000 & \cite{2005IBVS.5657....1H} \\
 & 2460135.57174	& 0.08059 & 24041.5 & -0.0024 & This study \\
 & 2460139.43157	& 0.06740 & 24055 & -0.0037 & This study \\
\hline
 & 2457944.979491 & & 0 & 0.00000 & VSX \\
 & 2457670.761095 & & -706 & 0.00274 & ASAS-SN \\
V0890 Lyr & 2459003.412275 & & 2725 & 0.00136 & \cite{2022BAVJ...60....1P} \\
 & 2460176.420936 & 0.000241 & 5745 & -0.00388 & This study \\
 & 2460177.392328 & 0.000442 & 5747.5 & -0.00352 & This study \\
\hline
\hline
\end{tabular}
\end{center}
\label{tab1}
\end{table*}

\begin{figure*}
    \centering
    \includegraphics[width=\textwidth]{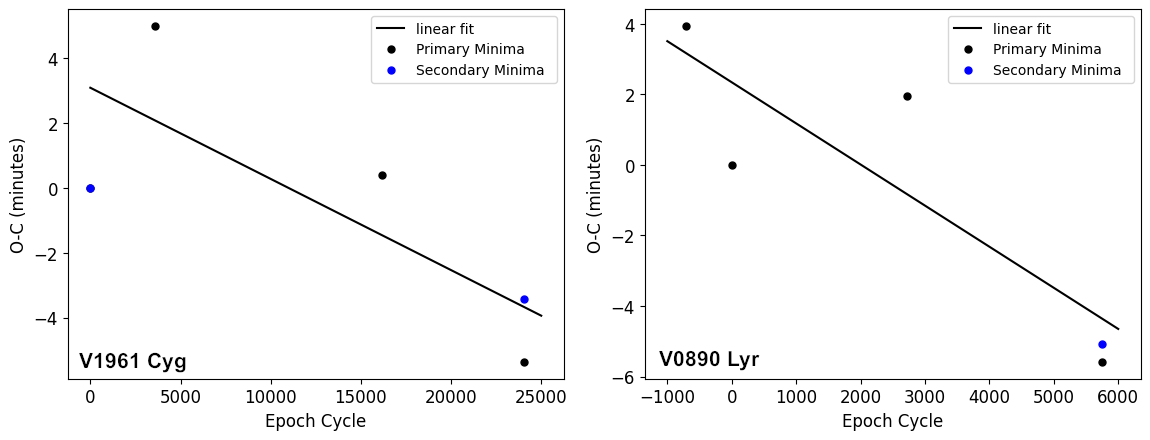}
    \caption{The left panel displays the O-C diagram for V1961 Cyg, while the right panel shows the O-C diagram for V0890 Lyr.}
    \label{OC}
\end{figure*}

A reference ephemeris was used to calculate the epoch and O-C values of each system. We chose $t_0(BJD_{TDB})=2453259.38056\pm0.00240$ from the \cite{2005IBVS.5657....1H} study and $P_0=0.2860135$ days from the ASAS-SN catalog for the V1961 Cyg system.
V0890 Lyr's $t_0(BJD_{TDB})=2457944.979491$ and $P_0=0.3884152$ days came from the VSX database. O-C diagrams of the systems are presented in Figure \ref{OC}. The O-C diagram of each system was fitted linearly based on the few observations available for these target systems.

We computed new ephemerides for these system (Equations \ref{eq:1} and \ref{eq:2}):

\begin{equation}\label{eq:1}
    \left\{
        \begin{array}{l}
            V1961\ Cyg:\\
            BJD_{TDB}(Min.I)=2453259.38270(184)+0.28601330(13)\times{E}
        \end{array}
    \right.
\end{equation}

\begin{equation}\label{eq:2}
    \left\{
        \begin{array}{l}
            V0890\ Lyr:\\
            BJD_{TDB}(Min.I)=2457944.98112(121)+0.3884144 (4)\times{E}
        \end{array}
    \right.
\end{equation}

where $E$ is the number of orbital cycles after the reference mid-eclipse time.

\vspace{1cm}
\section{Light Curve Solutions}
We employed the PHOEBE 2.4.9 Python code and the MCMC method in this investigation (\citealt{2016ApJS..227...29P}, \citealt{conroy2020physics}) to provide the first light curve analysis of the binary systems V1961 Cyg and V0890 Lyr. We have chosen a contact mode in the PHOEBE code based on the classification given to both systems in the catalogs and the typical shape of their light curves. The assumed gravity-darkening coefficients and bolometric albedo in this study are respectively $g_h=g_c=0.32$ (\citealt{lucy1967gravity}) and $A_h=A_c=0.5$ (\citealt{rucinski1969proximity}). A stellar atmosphere model has been applied under the \cite{castelli2004missing} method, and limb-darkening coefficients were determined by using PHOEBE.

We considered the Gaia DR3\footnote{\url{https://gea.esac.esa.int/archive/}} reported temperatures (5386 K) on the hotter component of V1961 Cyg as an initial value based on the light curve morphology. TESS reported a temperature for V1961 Cyg to be $5157\pm224$ K.
For the V0890 Lyr system, we set the effective temperature provided by the TESS database (6876 K) on the hotter component. Gaia DR3 did not report the V0890 Lyr system's temperature, while Gaia DR2 indicates it as 6911 K.
Then, we used the depth difference between the light curve's primary and secondary minima to determine the temperature ratio of each system.

\begin{figure*}
    \centering
    \includegraphics[width=\textwidth]{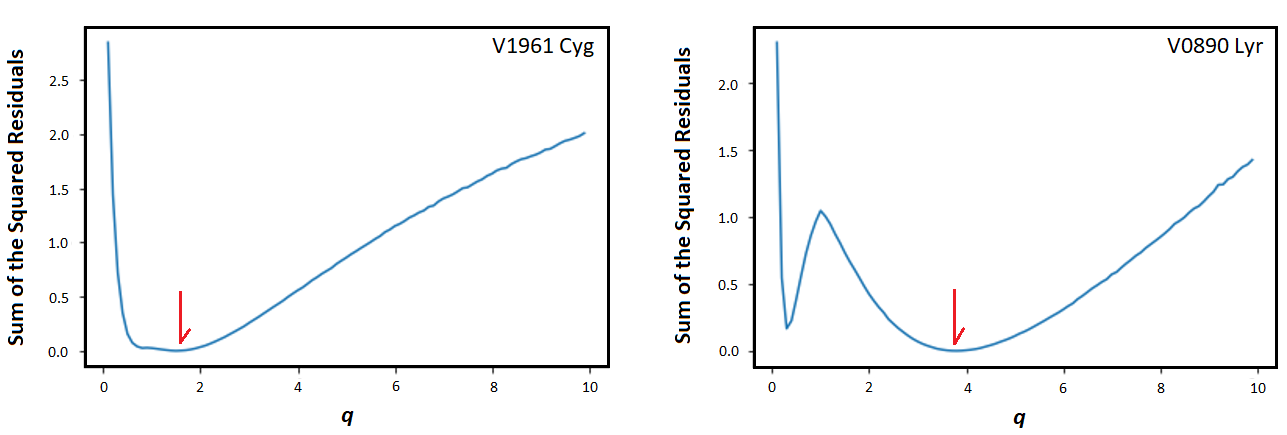}
    \caption{Sum of the squared residuals as a function of the mass ratio.}
    \label{q-search}
\end{figure*}

We determined the initial value of the mass ratio using photometric data. So, we performed a $q$-search with a 0.1 step, ranging from 0.1 to 10. Figure \ref{q-search} illustrates that each $q$-search curve has a clear minimum sum of squared residuals. Then, we tried to obtain a good initial theoretical fit to the observational data.

\begin{figure*}
    \centering
    \includegraphics[width=\textwidth]{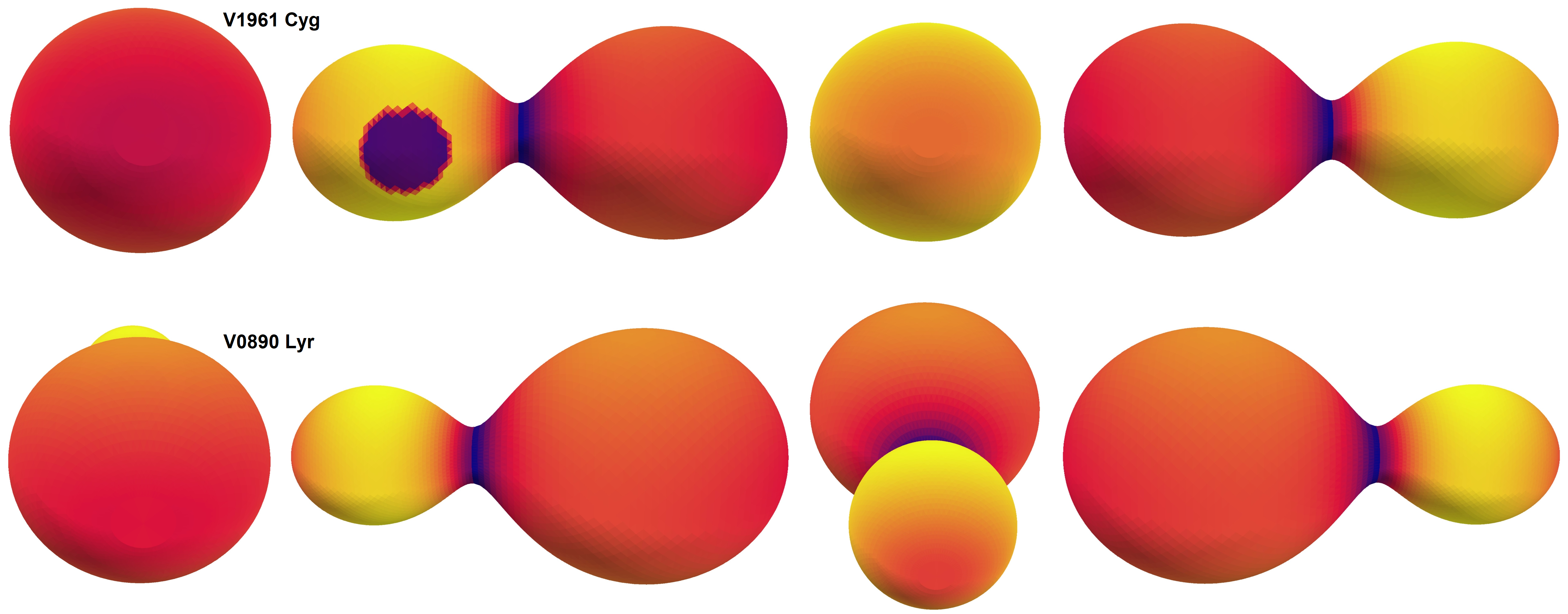}
    \caption{3D representation for V1961 Cyg (up) and V0890 Lyr (down). The displayed phases are 0, 0.25, 0.5, and 0.75.}
    \label{3D}
\end{figure*}

The light curve of the V1961 Cyg contact binary system shows asymmetry in the peak brightness. The difference between the maxima of the light curve for V1961 Cyg is $0.068^{mag}$. This is a sign of the O'Connell effect, for which one explanation might be the existence of a starspot due to magnetic activity on the star's surface (\citealt{1951MNRAS.111..642O}). Accordingly, one cold starspot has been added to the hotter component of the V1961 Cyg system. This starspot position and characteristic is depicted in Table \ref{tab2}, and it is visible in the 3D representation of V1961 Cyg in Figure \ref{3D}.

Then, we used PHOEBE's optimization tool to improve the theoretical fit. Finally, we used the MCMC approach, which is based on the emcee package (\citealt{2013PASP..125..306F}) in PHOEBE, to determine the values of the parameters together with their uncertainty (\citealt{hogg2018data}).
This process took into account a normal Gaussian distribution within the range of solutions for inclination ($i$), the mass ratio ($q$), fillout factor ($f$), effective temperatures for both components ($T_{1,2}$), and the luminosity of the primary star ($l_1$). Four starspot parameters were also added for the V1961 Cyg system's MCMC process. The MCMC approach was employed with 96 walkers and 1500 iterations for the V1961 Cyg, and 96 walkers and 1000 iterations for the V0890 Lyr.

\begin{figure*}
    \centering
    \includegraphics[width=\textwidth]{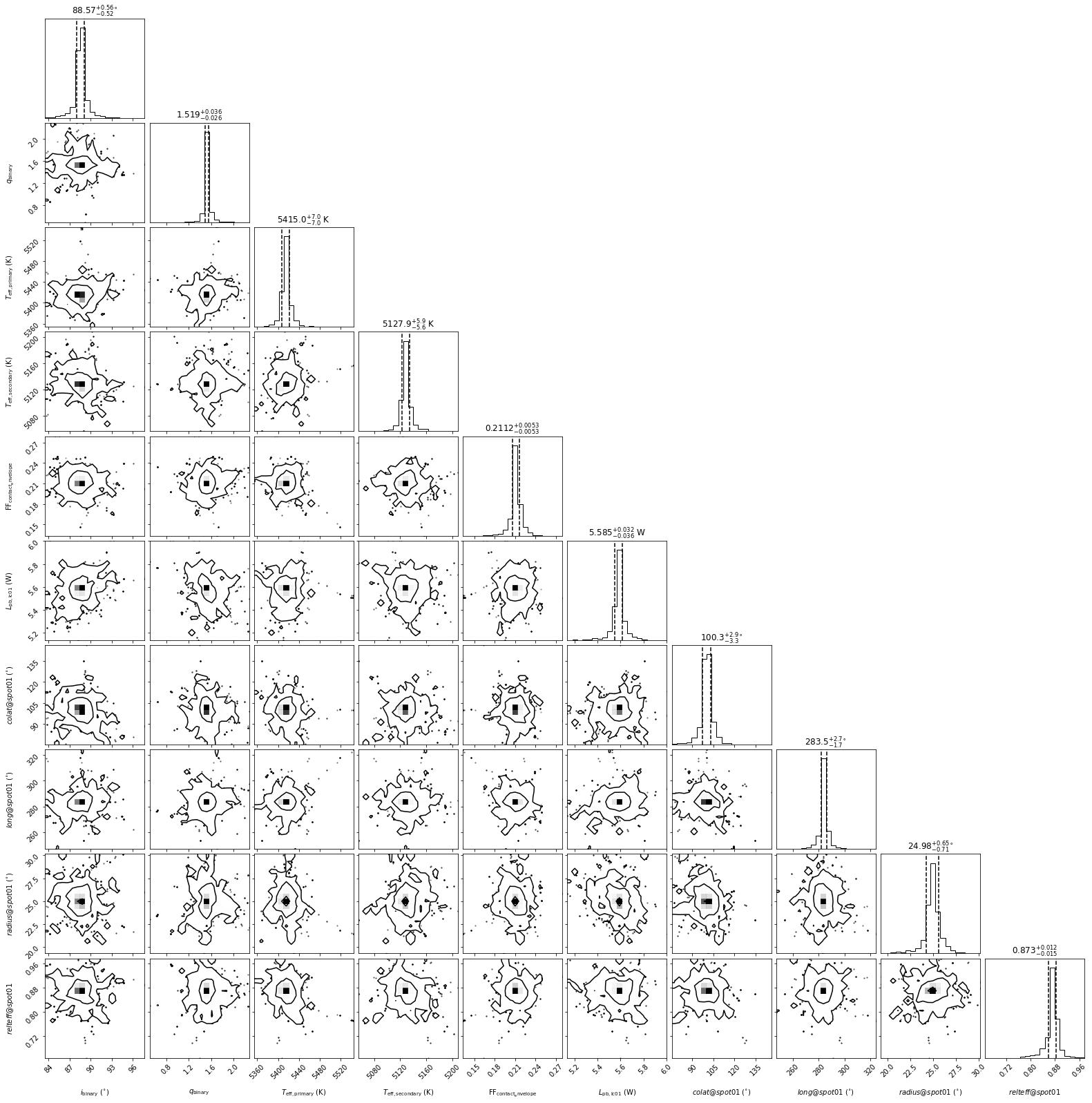}
    \caption{Determined corner plots of V1961 Cyg by the MCMC processing.}
    \label{Cornerplots1961}
\end{figure*}

\begin{figure*}
    \centering
    \includegraphics[width=\textwidth]{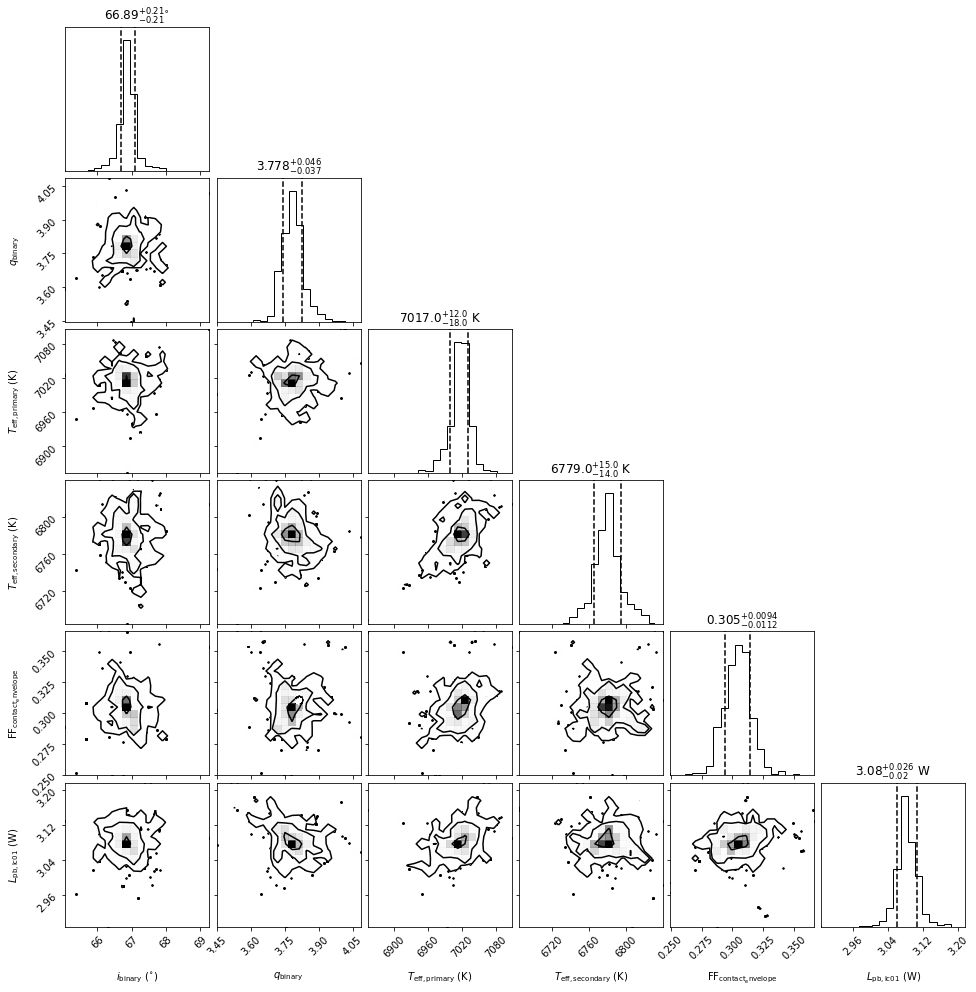}
    \caption{Determined corner plots of V0890 Lyr by the MCMC processing.}
    \label{Cornerplots890}
\end{figure*}

\begin{figure*}
    \centering
    \includegraphics[scale=0.36]{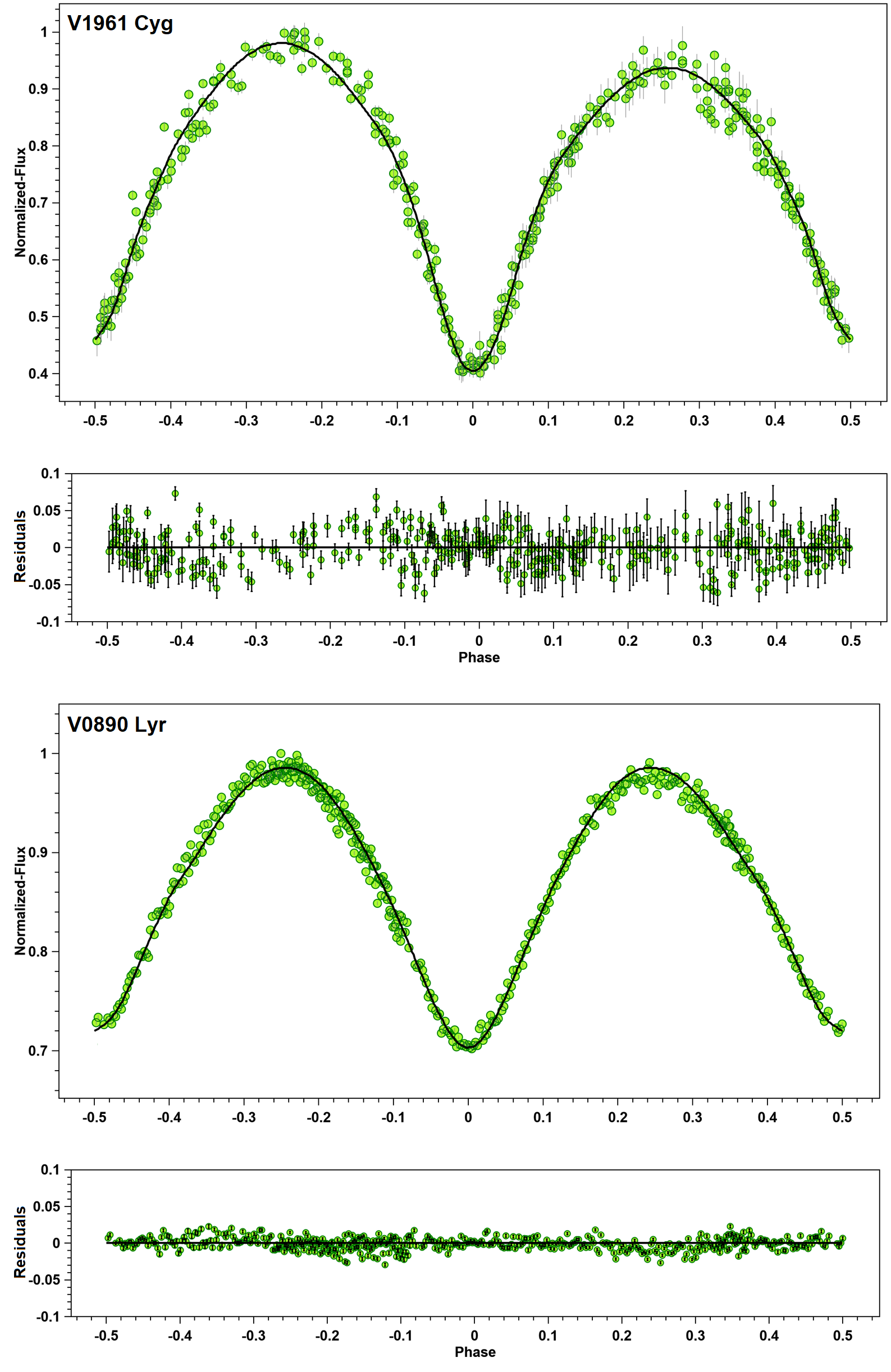}
    \caption{Light curves from observational data (green dots) and the best theoretical fit (black line) are presented for V1961 Cyg and V0890 Lyr. Residuals are displayed on the bottom panel of each light curve.}
    \label{LC}
\end{figure*}

The photometric light curve solutions' results are listed in Table \ref{tab2} and the corner plots produced by the MCMC modeling for each system are shown in Figures \ref{Cornerplots1961} and \ref{Cornerplots890}. The observational and synthetic light curves for the V1961 Cyg and V0890 Lyr systems are shown in Figure \ref{LC}.

\begin{table*}
\caption{Photometric light curve solutions' results for V1961 Cyg and V0890 Lyr.}
\centering
\begin{center}
\footnotesize
\begin{tabular}{c c c c c}
 \hline
 \hline
Parameter && V1961 Cyg && V0890 Lyr\\
\hline
$T_{h}$ (K) && $5415_{\rm-(7)}^{+(7)}$ && $7017_{\rm-(18)}^{+(12)}$\\
\\
$T_{c}$ (K) && $5128_{\rm-(6)}^{+(6)}$ && $6779_{\rm-(14)}^{+(15)}$\\
\\
$q=M_c/M_h$ && $1.519_{\rm-(26)}^{+(36)}$ && $3.778_{\rm-(27)}^{+(46)}$\\
\\
$\Omega_h=\Omega_c$ && $4.432(9)$ && $7.437(12)$\\
\\
$i^{\circ}$ &&	$88.57_{\rm-(52)}^{+(56)}$ && $66.89_{\rm-(21)}^{+(21)}$\\
\\
$f$ && $0.211_{\rm-(5)}^{+(5)}$ && $0.305_{\rm-(11)}^{+(9)}$\\
\\
$l_h/l_{tot}(V)$ && $0.482_{\rm-(4)}^{+(4)}$ && $0.264_{\rm-(4)}^{+(4)}$\\
\\
$l_c/l_{tot}(V)$ && $0.518(3)$ && $0.736(5)$\\
\\
$r_{(mean)h}$ && $0.362(7)$ && $0.289(7)$\\
\\
$r_{(mean)c}$ && $0.434(8)$ && $0.515(8)$\\
\\
Phase shift && $-0.01(1)$ && $-0.01(1)$\\
\hline
$Colatitude_{spot}(deg)$ && $100_{\rm-(3)}^{+(3)}$ &&\\
\\
$Longitude_{spot}(deg)$ && $284_{\rm-(2)}^{+(3)}$ &&\\
\\
$Radius_{spot}(deg)$ && $24.98_{\rm-(71)}^{+(65)}$ &&\\
\\
$T_{spot}/T_{star}$ && $0.873_{\rm-(15)}^{+(12)}$ &&\\
\\
Component && Hotter &&\\
\hline
\hline
\end{tabular}
\end{center}
\label{tab2}
\end{table*}

\vspace{1cm}
\section{Absolute Parameters}
The empirical relationship between the semi-major axis ($a$) and orbital period ($P$) from the \cite{2024PASP..136b4201P} study was utilized as the foundation for estimating the absolute parameters of each target system (Equation \ref{eq:3}).

\begin{equation}\label{eq:3}
a=(0.372_{\rm-0.114}^{+0.113})+(5.914_{\rm-0.298}^{+0.272})\times{P}
\end{equation}

We presented the orbital period of the systems as 0.28601330(13) days for V1961 Cyg and 0.3884144(4) days for V890 Lyr, respectively. Using these values, we have computed $a(R_{\odot})$ from equation \ref{eq:3}.

Equation \ref{eq:4} was used to estimate the radii of stars based on the $r_{mean(h,c)}$ which resulted in the light curve solutions, and the value of $a$.

\begin{equation}\label{eq:4}
R_{(h,c)}={a}\times{r_{mean(h,c)}}
\end{equation}

The effective temperature of each component has also been found in the light curve solutions. Considering a blackbody emission, we computed the luminosity $L_{(h,c)}(L_{\odot})$ from the radius of each component and its effective temperature (Equation \ref{eq:5}).

\begin{equation}\label{eq:5}
{L_{(h,c)}}=4\pi\sigma {T_{(h,c)}}^4 {R_{(h,c)}}^2
\end{equation}

The mass ratio and well-known Kepler's third law were used to estimate each component's mass. The stars mass values were computed by equations \ref{eq:6} and \ref{eq:7}.

\begin{eqnarray}
M{_h}=\frac{4\pi^2a^3}{GP^2(1+q)}\label{eq:6}\\
M{_c}=q\times{M{_h}}\label{eq:7}
\end{eqnarray}

Next, the surface gravity of each star is determined using its mass and radius by equation \ref{eq:8}.

\begin{equation}\label{eq:8}
\log{\left ( g{_{(h,c)}}\right )}=\log{\left (\frac{GM{_{(h,c)}}}{R_{(h,c)}^2}\right )}
\end{equation}

Finally, using the \cite{10.1093/mnras/17.1.12} study, we were able to compute the absolute bolometric magnitudes ($M_{bol(h,c)}$) of each component (Equation \ref{eq:9}). We used the solar bolometric magnitude value reported from the \cite{torres2010use} study, which is $M_{bol\odot}=4.73^{mag}$.

\begin{equation}\label{eq:9}
M_{bol(h,c)}=M_{bol\odot}-2.5\times log\left(\frac{L_{(h,c)}}{L_\odot}\right)
\end{equation}

Table \ref{tab3} lists the values of the estimated absolute parameters. The uncertainties of each absolute parameter in this study are computed using the $a-P$ relationship results and the light curve solutions parameters (${r_{mean(h,c)}}$, $T_{(h,c)}$, and $q$).

\begin{table*}
\caption{Absolute parameter estimation results.}
\centering
\begin{center}
\footnotesize
\begin{tabular}{c | c c | c c}
 \hline
 \hline
 & \multicolumn{2}{c|}{V1961 Cyg} & \multicolumn{2}{c}{V0890 Lyr}\\
Parameter & Hotter Star & Cooler Star & Hotter Star & Cooler Star\\
\hline
$M(M_\odot)$ & $0.572(178)$ & $0.870(294)$ & $0.354(93)$ & $1.338(369)$\\

$R(R_\odot)$ & $0.747(86)$ & $0.896(103)$ & $0.771(85)$ & $1.375(139)$\\

$L(L_\odot)$ & $0.432(109)$ & $0.500(124)$ & $1.293(315)$ & $3.582(748)$\\

$M_{bol}(mag.)$ & $5.641(243)$ & $5.483(241)$ & $4.451(237)$ & $3.345(206)$\\

$log(g)(cgs)$ & $4.449(23)$ & $4.473(32)$ & $4.213(10)$ & $4.288(27)$\\

$a(R_\odot)$ &  \multicolumn{2}{c|}{$2.063(195)$} & \multicolumn{2}{c}{$2.669(224)$}\\
\hline
\hline
\end{tabular}
\end{center}
\label{tab3}
\end{table*}

\vspace{1cm}
\section{Discussion and Conclusion}
We employed ground-based photometric observations to investigate two W UMa-type contact binary systems. V1961 Cyg and V0890 Lyr were observed during five nights in 2023 at an observatory in France. We extracted four primary and secondary times of minima from our data for both target systems and presented new ephemerides.
\\
\\
The first light curve analysis of the systems was done using PHOEBE and MCMC. We used the $q$-search method on photometric data to determine the initial mass ratio. The final parameter and uncertainty results were obtained after performing the MCMC process. Therefore, the companion temperature difference is 287 K and 238 K for V1961 Cyg and V0890 Lyr, respectively. The spectral types of V1961 Cyg stars and V0890 Lyr stars were determined to be G8-K1 and F1-F2, respectively, based on the \cite{eker2018interrelated} and \cite{cox2015allen} studies.

\begin{figure*}
    \centering
    \includegraphics[width=\textwidth]{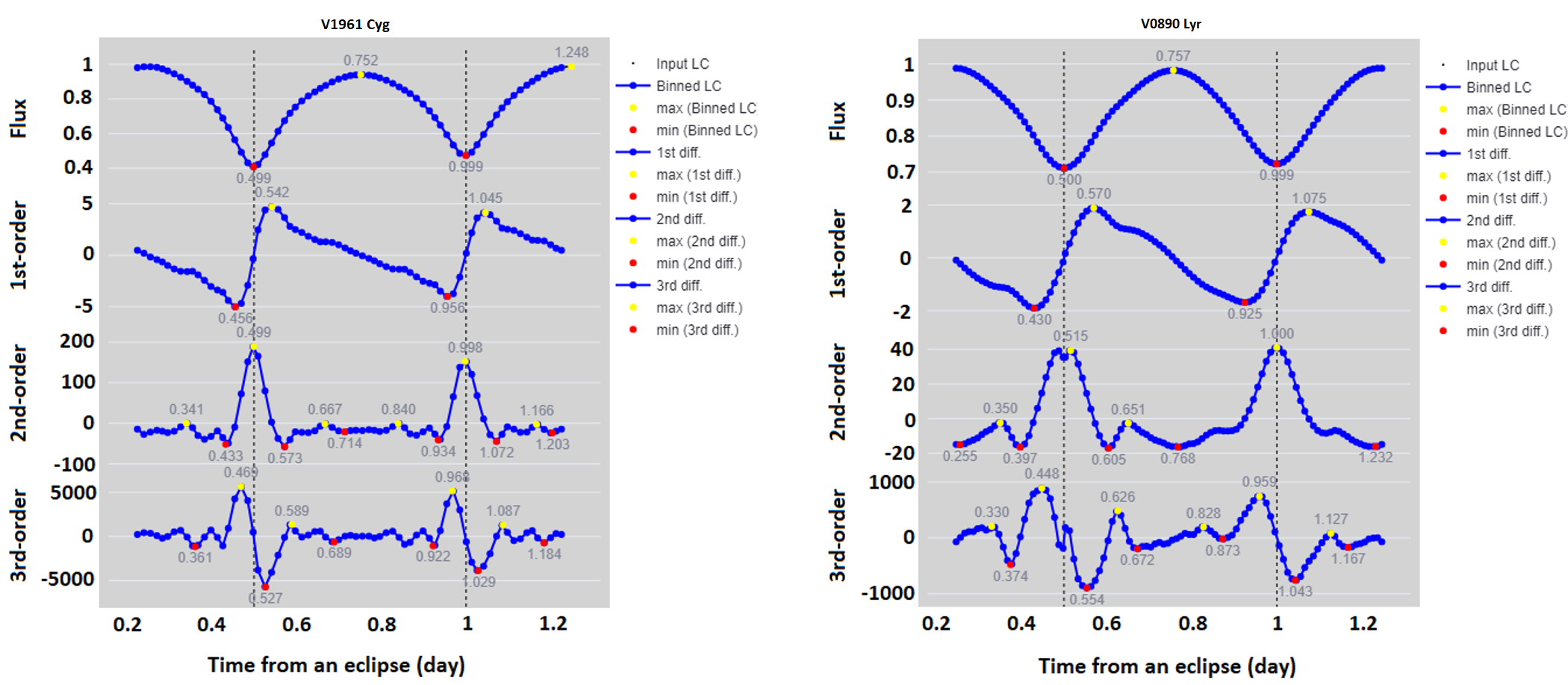}
    \caption{The light curve of V1961 Cyg (left) and V0890 Lyr (right), and first to third derivatives (top to bottom panels), respectively. The units of the panels on the vertical axis from top to bottom are W m$^{-2}$, 10 W m$^{-2}$ day$^{-1}$, $10^2$ W m$^{-2}$ day$^{-2}$, and $10^4$ W m$^{-2}$ day$^{-3}$, respectively.}
    \label{NEWq}
\end{figure*}

\begin{figure*}
    \centering
    \includegraphics[width=\textwidth]{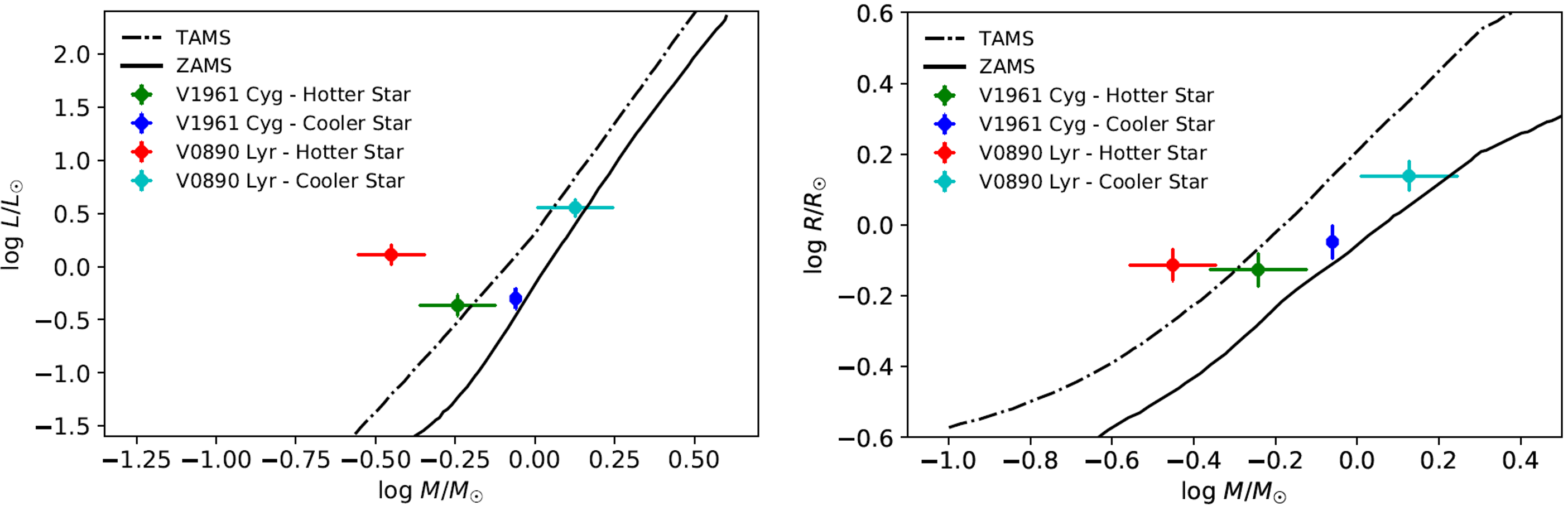}
    \caption{Both systems are displayed in a Mass-Luminosity diagram (left) and a Mass-Radius diagram (right) along with the ZAMS and TAMS limits.}
    \label{M-L-R}
\end{figure*}

We used a new method to examine the mass ratios resulting from our light curve solutions. \cite{2023ApJ...958...84K} presented a method for estimating the photometric mass ratio of overcontact binaries using derivatives of the light curve. Most overcontact systems can employ this new method. The basis of this method is derivation at various orders of the photometric light curve. The light curve should show at least two maxima after the derivative. A parameter named $W$ is obtained after the third-order derivative. According to \cite{2023ApJ...958...84K}, there is a strong relationship between $W$ and $q$. We used this method for V1961 Cyg and V0890 Lyr, and the results were mass ratios of $1/q=0.681(72)$ and $1/q=0.268(43)$, respectively. We found close mass ratio estimates with a discrepancy of $\Delta q=0.023$ and $\Delta q=0.003$ for V1961 Cyg and V0890 Lyr, respectively, between our light curve analysis results and the \cite{2023ApJ...958...84K} study. Figure \ref{NEWq} shows the derivative process of the light curves.
\\
\\
The estimation of the absolute parameters was presented using the relationship between the semi-major axis and the orbital period.
Based on the estimated absolute parameters, we displayed the state of evolution of the target systems on the Mass-Luminosity $(M-L)$ and Mass-Radius $(M-R)$ diagrams. The theoretical Zero-Age Main Sequence (ZAMS) and Terminal-Age Main Sequence (TAMS) lines are also displayed based on the study of \cite{2000AAS..141..371G} on the $M-L$ and $M-R$ diagrams (Figure \ref{M-L-R}). In the $M-L$ and $M-R$ diagrams, the hotter star of V1961 Cyg is near the TAMS line, whereas for V0890 Lyr, it is above the TAMS. The cooler star is located between ZAMS and TAMS for both target systems.

The orbital angular momentum ($J_0$) was computed for V1961 Cyg and V0890 Lyr to be $J_0=51.56\pm0.20$ and $J_0=51.56\pm0.17$, respectively, based on the orbital period, mass ratio, and total mass of the systems. We estimated $J_0$ using Equation \ref{eq10} that the \cite{2006MNRAS.373.1483E} study provided.

\begin{equation}\label{eq10}
J_0=\frac{q}{(1+q)^2} \sqrt[3] {\frac{G^2}{2\pi}M^5P}
\end{equation}

\begin{figure}
    \centering
    \includegraphics[scale=0.42]{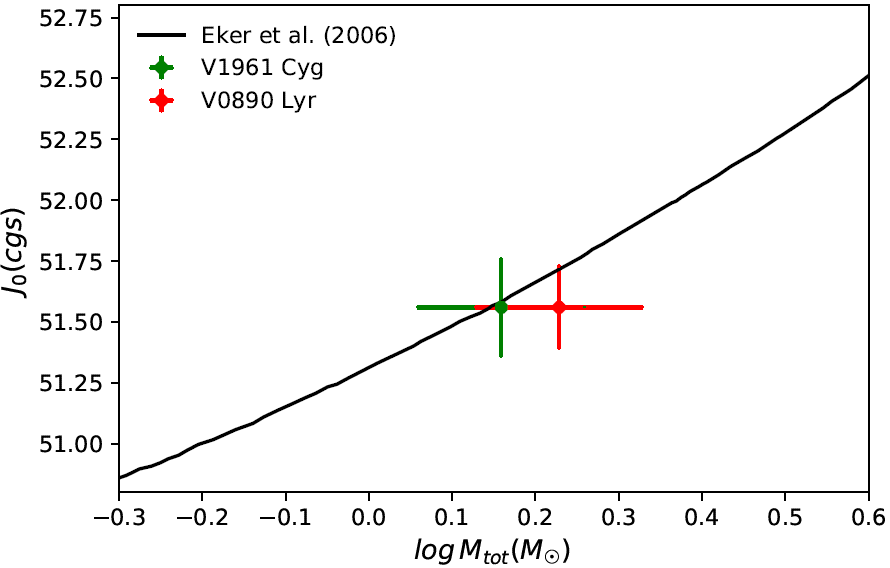}
    \caption{Diagram of orbital angular momentum of each system as a function of total mass.}
    \label{J0}
\end{figure}

Furthermore, the $logM_{tot}-logJ_0$ diagram (Figure \ref{J0}) shows the system's position and indicates that both systems are located in the region of contact binary systems.
\\
\\
In the literature, there are studies about the empirical relationship between the mass ratio ($q$) and the luminosity ratio ($L_{ratio}$) for contact binary systems. We have listed some of the relationships presented in these studies in Table \ref{tab4}.
We updated the $q-L_{ratio}$ relationship based on a valid sample.

\begin{table}
\caption{Some of the $q-L_{ratio}$ relationships presented by previous investigations for contact binary systems.}
\centering
\begin{center}
\footnotesize
\begin{tabular}{c c}
 \hline
 \hline
Relationship & References\\
\hline
$L_2/L_1=(M_2/M_1)^{0.92}$	&\cite{1968ApJ...153..877L}\\
$L_2/L_1=(M_2/M_1)^{0.96}$	&\cite{1968ApJ...151.1123L}\\
$L_2/L_1=(M_2/M_1)^{0.92}$	&\cite{1970AA.....5...12C}\\
$L_2/L_1=(M_2/M_1)^{0.93}$	&\cite{1981AAS...46..193H}\\
$L_2/L_1=(M_2/M_1)^{0.82}$	&\cite{1992ApSS.189..237R}\\
$L_2/L_1=(M_2/M_1)^{1.04}$	&\cite{1992ApSS.189..237R}\\
$L_2/L_1=(M_2/M_1)^{0.74}$	&\cite{2005JKAS...38...43A}\\
$L_2/L_1=(M_2/M_1)^{0.72}$ & \cite{2024RAA....24a5002P}\\
\hline
\hline
\end{tabular}
\end{center}
\label{tab4}
\end{table}

We chose a sample for the $q-L_{ratio}$ relationship from the \cite{2021ApJS..254...10L} study. Therefore, we only selected studies that had or used spectroscopic results. We excluded from the sample the results that contained $l_3$, or if $l_3$ in subsequent studies was present for that system. Also, we are not considering the studies that used low-resolution spectroscopic observations. The systems used for this sample include both subtypes A and W. As a result, we have listed the remaining 88 contact binary systems for the sample in Table \ref{tab5}. We used a linear fit on the points, whose equation is as follows:

\begin{equation}\label{eq11}
q=(1.066\pm0.052)L_{ratio}+(0.042\pm0.019)
\end{equation}

\begin{figure}
    \centering
    \includegraphics[scale=0.48]{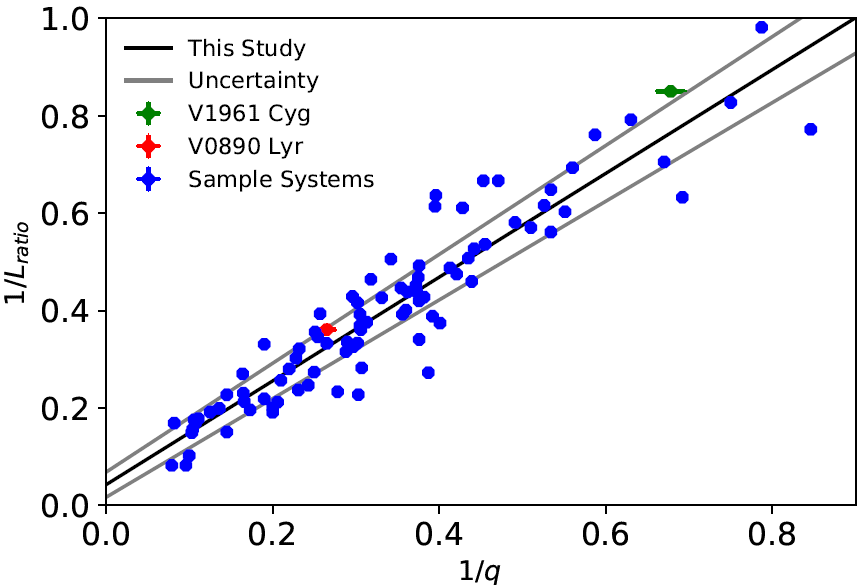}
    \caption{$q-L_{ratio}$ diagram using 88 contact binary systems and a linear fit on the points.}
    \label{q-Lr}
\end{figure}

The mass ratio and luminosity ratio in equation \ref{eq11} are expressed as $1/q$ and $1/L_{ratio}$, respectively, and have not been utilized in previous studies of similar form (Table \ref{tab4}). It seems that in this form, there is less scattering in the data points than in other samples. Additionally, mass ratios from both kinds of photometric and spectroscopic data can be found in other literature samples. The position of the V1961 Cyg and V0890 Lyr systems in Figure \ref{q-Lr} is in agreement with other contact systems that have spectroscopic data analyzed.
\\
\\
We used the $i>arccos|(r_1-r_2)/a|$ relationship from the \cite{2020ApJS..247...50S} study to find that the target systems are partial or total eclipse systems. Therefore, V1961 Cyg with $i=88.57^{\circ}$ is total, and V0890 Lyr with $i=67.00^{\circ}$ is a partial binary system. Based on the discussion of the \cite{2005ApSS.296..221T} study, systems with total and partial eclipses with high orbital inclinations can have acceptable photometric mass ratios. Also, according to the mass ratio from the light curve solutions in the MCMC process, the mass ratio results using the \cite{2023ApJ...958...84K} method, and their position in the empirical relationship of $q-L_{ratio}$, the obtained mass ratios are relabeled for both systems.

According to the mass ratios, fillout factors, and orbital inclinations of the target systems, we can conclude that V1961 Cyg and V0890 Lyr are contact binary systems. Based on the fillout factor, there are three categories: deep ($f\geq 50\%$), medium ($25\% \leq f < 50\%$), and shallow ($f<25\%$) eclipsing contact binary stars (\citealt{2022AJ....164..202L}). So, V1961 Cyg is shallow, and V0890 Lyr is a medium system. Additionally, contact binaries are divided into A- and W-subtypes (\citealt{binnendijk1970orbital}). The more massive component is a hotter star in the A-subtype, and if the less massive component has a higher effective temperature, it is classified as a W-subtype. Therefore, based on the results of the light curve solutions and the estimation of absolute parameters, they are categorized as a W-subtype since the less massive component has higher effective temperatures.

\begin{table*}
\caption{The sample used for the relationship between the mass ratio and the luminosity ratio.}
\centering
\begin{center}
\footnotesize
\begin{tabular}{c c c c c c c c}
\hline
\hline
System & $1/q$ & $L_{ratio}$ & Reference & System & $1/q$ & $L_{ratio}$ & Reference\\
\hline
AA UMa	&	0.551	&	1.177	&	\cite{2011PASP..123...34L}	&	NN Vir	&	0.491	&	1.021	&	\cite{2011MNRAS.412.1787D}	\\
AB And	&	0.560	&	1.687	&	\cite{2014NewA...30...64L}	&	NSVS 4161544	&	0.296	&	0.842	&	\cite{2019AJ....157...73K}	\\
AD Phe	&	0.376	&	0.990	&	\cite{2017AJ....154..260P}	&	NSVS 13392702	&	0.510	&	1.961	&	\cite{2014AA...563A..34L}	\\
AH Aur	&	0.165	&	0.334	&	\cite{2005AcA....55..123G}	&	OO Aql	&	0.846	&	1.669	&	\cite{2016RAA....16....2L}	\\
AP Leo	&	0.297	&	0.690	&	\cite{2003AA...412..465K}	&	OU Ser	&	0.173	&	0.583	&	\cite{2011MNRAS.412.1787D}	\\
AQ Tuc	&	0.354	&	0.595	&	\cite{2001CoSka..31....5C}	&	QX And	&	0.306	&	0.742	&	\cite{2011AA...525A..66D}	\\
AU Ser	&	0.692	&	1.790	&	\cite{2018IBVS.6256....1A}	&	RT LMi	&	0.382	&	1.019	&	\cite{2010MNRAS.408..464Z}	\\
BB Peg	&	0.318	&	0.880	&	\cite{2020NewA...7801354K}	&	RW Com	&	0.471	&	1.984	&	\cite{2011MNRAS.412.1787D}	\\
BI CVn	&	0.413	&	1.075	&	\cite{2014NewA...29...57N}	&	RW Dor	&	0.630	&	2.207	&	\cite{2019PASJ...71...34S}	\\
BN Ari	&	0.392	&	1.309	&	\cite{2018AcA....68..159A}	&	RZ Tau	&	0.376	&	0.905	&	\cite{2003AJ....126.1960Y}	\\
BO Ari	&	0.190	&	0.597	&	\cite{2015NewA...39....9G}	&	SW Lac	&	0.787	&	2.454	&	\cite{2005AcA....55..123G}	\\
BX Dra	&	0.288	&	0.497	&	\cite{2013PASJ...65....1P}	&	SX Crv	&	0.079	&	0.250	&	\cite{2004AcA....54..299Z}	\\
BX Peg	&	0.376	&	1.341	&	\cite{2015NewA...41...17L}	&	TV Mus	&	0.166	&	0.372	&	\cite{2005AJ....130..224Q}	\\
CC Com	&	0.526	&	2.383	&	\cite{2011AN....332..626K}	&	TW Cet	&	0.750	&	2.367	&	\cite{2011MNRAS.412.1787D}	\\
CK Boo	&	0.111	&	0.313	&	\cite{2011MNRAS.412.1787D}	&	TX Cnc	&	0.455	&	1.188	&	\cite{2011MNRAS.412.1787D}	\\
CN And	&	0.387	&	0.836	&	\cite{2019RAA....19...10Y}	&	TY Pup	&	0.250	&	0.305	&	\cite{2018AJ....156..199S}	\\
DK Cyg	&	0.307	&	0.652	&	\cite{2015AJ....149..194L}	&	TY UMa	&	0.396	&	1.117	&	\cite{2015AJ....149..120L}	\\
DN Boo	&	0.103	&	0.230	&	\cite{2008NewA...13..468S}	&	U Peg	&	0.331	&	0.883	&	\cite{2001AA...367..840D}	\\
DN Cam	&	0.442	&	0.887	&	\cite{2004AcA....54..195B}	&	UV Lyn	&	0.372	&	0.896	&	\cite{2005AcA....55..389Z}	\\
DX Tuc	&	0.290	&	0.769	&	\cite{2007AA...465..943S}	&	UX Eri	&	0.373	&	0.838	&	\cite{2011MNRAS.412.1787D}	\\
DY Cet	&	0.356	&	0.808	&	\cite{2011MNRAS.412.1787D}	&	V1073 Cyg	&	0.303	&	0.386	&	\cite{2018RAA....18...20T}	\\
DZ Psc	&	0.136	&	0.371	&	\cite{2013AJ....146...35Y}	&	V1128 Tau	&	0.534	&	1.749	&	\cite{2014AJ....148..126C}	\\
EF Boo	&	0.531	&	1.773	&	\cite{paki2024reanalyzing}	&	V1191 Cyg	&	0.107	&	0.341	&	\cite{2014NewA...31...14O}	\\
EQ Tau	&	0.439	&	1.286	&	\cite{2015NewA...34..262H}	&	V1453 Her	&	0.670	&	2.958	&	\cite{2014AA...563A..34L}	\\
ET Leo	&	0.342	&	0.987	&	\cite{2006AcA....56..127G}	&	V1918 Cyg	&	0.278	&	0.673	&	\cite{2016NewA...47...57G}	\\
EX Leo	&	0.200	&	0.489	&	\cite{2010MNRAS.408..464Z}	&	V2357 Oph	&	0.231	&	0.556	&	\cite{2011MNRAS.412.1787D}	\\
EZ Hya	&	0.257	&	0.571	&	\cite{2004PASP..116..826Y}	&	V2377 Oph	&	0.395	&	0.929	&	\cite{2011MNRAS.412.1787D}	\\
FG Hya	&	0.104	&	0.317	&	\cite{2010MNRAS.408..464Z}	&	V351 Peg	&	0.360	&	0.607	&	\cite{2005NewA...10..163A}	\\
FP Boo	&	0.096	&	0.150	&	\cite{2006AcA....56..127G}	&	V357 Peg	&	0.401	&	0.693	&	\cite{2012NewA...17..603E}	\\
FU Dra	&	0.251	&	0.818	&	\cite{2012PASJ...64...48L}	&	V402 Aur	&	0.200	&	0.331	&	\cite{2004AcA....54..299Z}	\\
GM Dra	&	0.210	&	0.620	&	\cite{2005AcA....55..123G}	&	V404 Peg	&	0.243	&	0.580	&	\cite{2011AN....332..690G}	\\
GR Vir	&	0.106	&	0.306	&	\cite{2005AcA....55..123G}	&	V417 Aql	&	0.362	&	0.978	&	\cite{2011MNRAS.412.1787D}	\\
GSC 03334-00553	&	0.421	&	1.118	&	\cite{2019AJ....157...73K}	&	V535 Ara	&	0.302	&	0.480	&	\cite{2012NewA...17..143O}	\\
GW Cnc	&	0.265	&	0.942	&	\cite{2016NewA...46...31G}	&	V546 And	&	0.254	&	0.663	&	\cite{2015NewA...36..100G}	\\
HH Boo	&	0.587	&	1.842	&	\cite{2015NewA...41...26G}	&	V781 Tau	&	0.453	&	1.313	&	\cite{2016ApSS.361...63L}	\\
HI Pup	&	0.206	&	0.476	&	\cite{2014NewA...31...56U}	&	V829 Her	&	0.435	&	1.215	&	\cite{2006NewA...12..192E}	\\
HN UMa	&	0.145	&	0.379	&	\cite{2007ASPC..362...82O}	&	V839 Oph	&	0.305	&	0.746	&	\cite{2011MNRAS.412.1787D}	\\
HV Aqr	&	0.145	&	0.387	&	\cite{2013NewA...21...46L}	&	V870 Ara	&	0.082	&	0.205	&	\cite{2007AA...465..943S}	\\
HV UMa	&	0.190	&	0.535	&	\cite{2000AA...356..603C}	&	V972 Her	&	0.164	&	0.370	&	\cite{2018ApSS.363...34S}	\\
KIC 10618253	&	0.125	&	0.286	&	\cite{2016PASA...33...43S}	&	VW Boo	&	0.428	&	1.250	&	\cite{2011AJ....141..147L}	\\
KIC 9832227	&	0.228	&	0.498	&	\cite{2017ApJ...840....1M}	&	VW Cep	&	0.302	&	1.085	&	\cite{2018AA...612A..91M}	\\
LO And	&	0.305	&	0.802	&	\cite{2015IBVS.6134....1N}	&	VY Sex	&	0.313	&	0.706	&	\cite{2011MNRAS.412.1787D}	\\
LS Del	&	0.375	&	1.031	&	\cite{2011MNRAS.412.1787D}	&	XX Sex	&	0.100	&	0.185	&	\cite{2011MNRAS.412.1787D}	\\
MW Pav	&	0.220	&	0.277	&	\cite{2015PASP..127..742A}	&	YY CrB	&	0.232	&	0.616	&	\cite{2005AcA....55..123G}	\\
\hline
\hline
\end{tabular}
\end{center}
\label{tab5}
\end{table*}

\vspace{1cm}
\section*{Data Availability}
Ground-based data will be made available on request.

\vspace{2cm}
\section*{Acknowledgements}
This manuscript was prepared by the BSN (\url{https://bsnp.info/}) project. We have made use of data from the European Space Agency (ESA) mission Gaia (\url{http://www.cosmos.esa.int/gaia}), processed by the Gaia Data Processing and Analysis Consortium (DPAC).

\vspace{1cm}
\section*{ORCID iDs}
\noindent Sabrina Baudart: 0009-0004-8426-4114\\
Atila Poro: 0000-0002-0196-9732\\

\vspace{1cm}
\bibliography{Ref}{}
\bibliographystyle{aasjournal}

\end{document}